\documentclass[runningheads,a4paper]{llncs}

\usepackage{amssymb, amsmath}
\usepackage{graphicx}
\usepackage[normalem]{ulem}
\usepackage{xspace}
\usepackage[usenames,dvipsnames]{xcolor}

\usepackage{url}
\urldef{\mailsa}\path|{meyerhenke,sanders, sebastian.schlag, christian.schulz}@kit.edu|  
\urldef{\mailsb}\path|vitali.henne@gmail.com|      
\newcommand{\keywords}[1]{\par\addvspace\baselineskip
\noindent\keywordname\enspace\ignorespaces#1}

\usepackage{multirow}
\usepackage[autolanguage]{numprint}

\usepackage{siunitx}

\usepackage{pdflscape}
\usepackage{longtable}
\usepackage{sidecap}

\usepackage[strict]{changepage}

\usepackage[hidelinks]{hyperref}

\usepackage[algo2e,ruled,vlined,linesnumbered]{algorithm2e}
\SetCommentSty{textit}
\DontPrintSemicolon
\IncMargin{-\parindent}
\SetAlCapHSkip{0pt}
\SetAlgoLined

\SetKwProg{Function}{Function}{}{}
\SetKwFunction{Activate}{Activate}
\SetKwFunction{rate}{rate}
\SetKwFunction{coarsen}{coarsen}
\SetKwFunction{DeltaGainUpdate}{DeltaGainUpdate}
\DeclareMathOperator*{\argmax}{\arg\!\max}

\newcommand{\Rem}[1]{\tcp*{#1}}
\newcommand{\Remi}[1]{\tcp*[f]{#1}}

\newcommand{\ie}{i.\,e.\xspace}
\newcommand{\eg}{e.\,g.\xspace}

\newcommand{\psan}[1]{{\color{red}[PS: #1]}}
\newcommand{\sschl}[1]{{\color{ForestGreen}[SS: #1]}}
\newcommand{\hmey}[1]{{\color{orange}[HM: #1]}}
\newcommand{\csch}[1]{{\color{blue}[CS: #1]}}
\renewcommand{\hmey}[1]{}
\renewcommand{\csch}[1]{}
\renewcommand{\sschl}[1]{}
\renewcommand{\psan}[1]{}

\begin{document}
\npthousandsep{\ }
\npdecimalsign{.} 
\mainmatter  

\title{$n$-Level Hypergraph Partitioning}

\titlerunning{$n$-Level Hypergraph Partitioning}

\author{Vitali Henne,
  Henning Meyerhenke,
  Peter Sanders,\\ 
  Sebastian Schlag,
  Christian Schulz}
\authorrunning{$n$-Level Hypergraph Partitioning}

\institute{Karlsruhe Institute of Technology (KIT), 76128 Karlsruhe, Germany\\
\mailsa \\
\mailsb}

%
%

\toctitle{$n$-Level Hypergraph Partitioning}
\tocauthor{Authors' Instructions}
\maketitle

\begin{abstract}
We develop a multilevel algorithm for hypergraph partitioning that contracts the vertices one at a time and thus allows very high quality. This includes a rating function that avoids nonuniform vertex weights, an efficient ``semi-dynamic'' hypergraph data structure, a very fast coarsening algorithm, and two new local search algorithms. One is a $k$-way hypergraph adaptation of Fiduccia-Mattheyses local search and gives high quality at reasonable cost. The other is an adaptation of size-constrained label propagation to hypergraphs. Comparisons with hMetis and PaToH indicate that the new algorithm yields better quality over several benchmark sets and has a running time that is comparable to hMetis. Using label propagation local search is several times faster than hMetis and gives better quality than PaToH for a VLSI benchmark set. 

\keywords{hypergraph partitioning, local search, label propagation}
\end{abstract}

\section{Introduction} \label{Introduction}
\paragraph{Context.}
Hypergraph partitioning (HGP) is an important problem with many application areas. Two prominent areas are 
VLSI design and scientific computing (\eg the acceleration of sparse matrix-vector multiplications)~\cite{Papa2007}.
While the former is an example of a field where small optimizations can lead to significant savings, the latter
is an example where hypergraph-based modeling better captures the objectives of the 
application domain~\cite{PaToH} than graph-based approaches. We focus on a version of the problem that partitions the vertices of a given
hypergraph into $k$ blocks of roughly equal size (in our case $1+\varepsilon$ times the average block size) while optimizing
an objective function. In this paper, we minimize the total cut size, i.e., the number of hyperedges that span multiple blocks.

Since the 1990s HGP has evolved into a broad research area~\cite{Alpert19951,DBLP:conf/dimacs/2012,Papa2007}. 
The two most widely used general-purpose tools are PaToH~\cite{PaToH} (originating from scientific computing) 
and hMetis~\cite{hMetisRB,hMetisKway} (originating from VLSI design). 
Other tools with certain distinguishing characteristics are known, in particular Mondriaan~\cite{Mondriaan} (matrix 
partitioning), MLPart~\cite{MLPart} (circuit partitioning), Zoltan~\cite{Zoltan} and Parkway~\cite{Parkway2.0} (parallel), 
and UMPa~\cite{DBLP:conf/dimacs/CatalyurekDKU12} (multi-objective).
All these tools use the \emph{multilevel paradigm}, which  has three phases. The first of which recursively \emph{coarsens} 
the hypergraph to obtain a hierarchy of smaller hypergraphs that reflect the basic structure of the input. After 
applying an \emph{initial partitioning} algorithm to the smallest hypergraph in the second phase, coarsening is 
undone and, at each level, a \emph{local search} method is used to improve the partition induced by the coarser level.

The two most popular local search approaches are greedy algorithms~\cite{hMetisKway,DBLP:conf/dimacs/CatalyurekDKU12} or variations of
the Fiduccia-Mattheyses (FM) heuristic~\cite{FM82}. FM-type algorithms move vertices to other 
blocks in the order of improvements in the objective. Since it allows to worsen the objective temporarily, FM can escape
local optima to some extent -- as opposed to simple greedy methods. However, currently only partitioners based on recursive bisection 
use FM-based local search algorithms~\cite{MLPart,PaToH,Zoltan,hMetisRB,Mondriaan}. On the other hand, direct $k$-way hypergraph 
partitioners~\cite{Aykanat:2008,hMetisKway,Parkway2.0,DBLP:conf/dimacs/CatalyurekDKU12} \emph{always} employ greedy methods, although generalizations of $2$-way FM to $k$-way partitioning have been proposed by Sanchis~\cite{HypergraphKFM}. 

When improving a $k$-way partition directly, each vertex can potentially be moved to $k-1$ other blocks. Sanchis's algorithm maintains these
moves in $k-1$ priority queues (PQs) for each block, resulting in $k(k-1)$ PQs in total.
Hence, previous work on multilevel HGP notes two reasons for resorting to greedy methods: 
(i) Working with $k(k-1)$ PQs limits the practicality to small values of $k$ and (ii) the Sanchis algorithm has been observed to be trapped early in local minima when used \emph{without} the multilevel framework~\cite{flatKFM,hMetisKway}.

\paragraph{Motivation and Contribution.}
To our knowledge, the reasons above have kept other partitioners from evaluating direct $k$-way local search
algorithms based on FM in the multilevel context. The present paper closes this gap with the following 
contributions, described in detail in \autoref{nHGP}:
(i) We present the first direct $k$-way $n$-level hypergraph partitioner. It is motivated by the success of $n$-level graph 
partitioning~\cite{nGP} that performs a very fine-grained coarsening by only contracting a single edge on each level of the multilevel hierarchy. 
(ii) Generalizing a greedy local search method based on size-constrained label propagation (SCLaP)~\cite{LPAgraphPartitioning}, we provide indication that greedy algorithms may work well in some cases, but cannot escape from relatively poor local optima in others.
(iii) We therefore propose a localized FM-based $k$-way local search algorithm along the lines of Sanchis~\cite{HypergraphKFM}
that is started with a single pair of vertices only.

On 164 out of 252 experiments on  well established benchmark sets, our FM-based $k$-way local search computes better partitions than both hMetis and PaToH and produces
 partitions of equal quality in 17 out of the 88 remaining cases (see \autoref{Experiments}).
Moreover, our algorithm is about as fast as hMetis. The  speed of our algorithm is mainly due to (i)
a semi-dynamic hypergraph data structure, (ii) engineering the coarsening phase with the aim of
uniform coarsening, and (iii)
employing various speed-up techniques~\cite{Caldwell,FM82,LockedNets} to accelerate
the gain update step, which is the main bottleneck of most FM implementations~\cite{Papa2007}. 

\vfill
\section{Preliminaries} \label{Preliminaries}
An \textit{undirected hypergraph} $H=(V,E,c,\omega)$ is defined as a set of vertices $V$ and a
set of hyperedges $E$ with vertex weights $c:V \rightarrow \mathbb{R}_{\geq0}$ and hyperedge 
weights $\omega:E \rightarrow \mathbb{R}_{>0}$, where each hyperedge is a subset of the vertex set $V$ (i.e., $e \subseteq V$).
We use $n$ to denote the number of hypernodes and $m$ to denote the number of hyperedges. In
HGP literature, hyperedges are also called \emph{nets} and the vertices of a net are called \emph{pins} \cite{PaToH}.
We extend $c$ and $\omega$ to sets, i.e., $c(U) :=\sum_{v\in U} c(v)$ and $\omega(F) :=\sum_{e \in F} \omega(e)$.
A vertex $v$ is \textit{incident} to a net $e$ if $ \{v\} \subseteq e$. We use $\mathrm{I}(v)$ to denote the set of all nets incident to a vertex $v$. 
The \textit{degree} $d(v)$ of a vertex $v$ is the number of its incident nets: $d(v) := |I(v)|$.
Two vertices are \textit{adjacent} if there exists a net $e$ that contains both vertices. The set $\Gamma(v) := \{ u~|~\exists~e \in E : \{v,u\} \subseteq e\}$ denotes of neighbors of $v$.
The \textit{size} $|e|$ of a net $e$ is the number of its pins. Nets of size one are called \emph{single-node} nets.

A \emph{$k$-way partition} of a hypergraph $H$ is a partition of its vertex set into $k$ \emph{blocks} $\mathrm{\Pi} = \{V_1, \dots, V_k\}$ 
such that $\bigcup_{i=1}^k V_i = V$, $V_i \neq \emptyset $ for $1 \leq i \leq k$ and $V_i \cap V_j = \emptyset$ for $i \neq j$. We use $b[v]$ to refer to the block id of vertex $v$.
We call a $k$-way partition $\mathrm{\Pi}$ \emph{$\mathrm{\varepsilon}$-balanced} if each block $V_i \in \mathrm{\Pi}$ satisfies a \emph{balance constraint}:
$\forall i \in \{1..k\}: |V_i| \leq L_{max} := (1+\varepsilon)\lceil \frac{|V|}{k} \rceil$ for some parameter $\mathrm{\varepsilon}$. 
We call a block $V_i$ \emph{overloaded} if $|V_i| > L_{max}$ and \emph{underloaded} if $|V_i| < L_{max}$.

Given a $k$-way partition $\mathrm{\Pi}$, the number of pins of a net $e$ in block $V_i$ is defined as 
$\mathrm{\Phi}(e,V_i) := |\{v \in V_i~|~v \in e \}|$. If $\mathrm{\Phi}(e, V_i) > 0$, we say that net $e$ \emph{is connected} to block $V_i$. Similarly, we say that a block $V_i$ is \emph{adjacent} to a vertex $v \notin V_i$ if $\exists~e \in  I(v): \mathrm{\Phi}(e, V_i) > 0$. $\mathrm{R}(v)$ denotes the set of all blocks adjacent to $v$.
For each net $e$, $\mathrm{\Lambda}(e) := \{V_i~|~ \mathrm{\Phi}(e, V_i) > 0\}$ denotes the \emph{connectivity set} of $e$. We define the \emph{connectivity} of a net $e$ as the cardinality of its connectivity set: $\mathrm{\lambda}(e) := |\mathrm{\Lambda}(e)|$ \cite{PaToH}.
We call a net  \emph{internal} if $\mathrm{\lambda}(e) = 1$ and \emph{cut} net otherwise (i.e., $\mathrm{\lambda}(e) > 1$). Analogously a vertex that is contained in at least one cut net is called \emph{border vertex}.

The \emph{$k$-way hypergraph partitioning problem} is to find an $\varepsilon$-balanced $k$-way partition of a hypergraph $H$ that minimizes the \emph{total cut} $ \omega(E')$ where $E' := \{e \in E : \mathrm{\lambda}(e) > 1\}$ for some $\varepsilon$. This problem is known to be NP-hard \cite{Lengauer:1990}.

\emph{Contracting} a pair of vertices $(u, v)$ means merging $v$ into $u$. The weight of $u$ becomes $c(u) := c(u) + c(v)$ and we 
connect $u$ to the former neighbors $\Gamma(v)$ of $v$. We refer to $u$ as the \emph{representative} and $v$ as the \emph{contraction partner}. 
This process can lead to parallel nets (i.e., $\exists~e_i,e_j \in E : e_i~\triangle~e_j \neq \emptyset$, where $\triangle$ is the 
symmetric difference). In this case, we choose net $e_i$ as representative, update its weight to $\omega(e_i) = \omega(e_i) + \omega(e_j)$ 
and remove $e_j$ from the hypergraph. If a contraction creates single-node nets we remove them from the hypergraph, since such nets 
can never become part of the cut. \emph{Uncontracting} a vertex $u$ undoes the contraction and restores removed parallel and single-node nets. The uncontracted vertex $v$ is put in the same 
block as $u$ and the weight of $u$ is set back to $c(u) := c(u) - c(v)$.
\vfill

\section{$n$-Level $k$-way Hypergraph Partitioning}  \label{nHGP}
We now present our main contributions. A high-level overview of our $n$-level hypergraph partitioning framework is provided in Algorithm~\ref{alg:nHGP}. As other multilevel algorithms our algorithm has  a coarsening, initial partitioning and an uncoarsening phase.
During the coarsening phase, we successively shrink the hypergraph by contracting only \emph{a single pair} of vertices \emph{at each level},
until it is small enough to be initially partitioned by some other partitioning algorithm. We desribe the details of our coarsening
algorithm in \autoref{Coarsening} and briefly discuss initial partitioning in \autoref{InitialPartitioning}. The initial solution is transfered
to the next finer level by performing a \emph{single} uncontraction step. Afterwards, one of our localized local search algorithms described in
\autoref{localizedFM} and \autoref{LPAlocalSearch} is used to further improve the solution quality.

\begin{algorithm2e}[b]
\caption{Multilevel Hypergraph Partitioning Framework}\label{alg:nHGP}\normalsize
\KwIn{Hypergraph $H$, number of desired blocks $k$, balance parameter $\varepsilon$.}
 
 \While(\Remi{coarsening phase})
    {$H$ is not small enough} {
     $(u,v) := \argmax_{u \in V} score(u)$ \Remi{choose vertex pair with highest rating}
      $H := \FuncSty{contract}(H,u,v)$ \Remi{$H := H \setminus \{v\}$}  
    }
  
  $ \mathrm{\Pi} := \FuncSty{partition}(H,k,\varepsilon)$ \Remi{initial partitioning phase}

  \While(\Remi{uncoarsening phase})
    {$H$ is not completely uncoarsened} {
    $(H, \mathrm{\Pi}, u,v) := \FuncSty{uncontract}(H,\mathrm{\Pi})$

    $(H, \mathrm{\Pi}) := \FuncSty{refine}(H, \mathrm{\Pi}, u, v, k, \varepsilon)$
    }
   \KwOut{$\varepsilon$-balanced $k$-way partition $\mathrm{\Pi}=\{V_1, \dots, V_k\}$}
\end{algorithm2e}

\paragraph{Hypergraph Data Structure.}
Traditional multilevel algorithms create a new hypergraph  for each level of the hierarchy. This is not feasible
in the $n$-level context, since storing each level explicitly would lead to quadratic space consumption. 
We therefore designed a \emph{semi-dynamic} hypergraph data structure that supports efficient contraction and uncontraction operations.
Conceptually, we represent the hypergraph $H$ as an undirected \emph{bipartite} graph $G=(W, F)$. The vertices and nets
of $H$ form the vertex set $W$. For each net $e$ incident to a vertex $v$, we add an edge $(e,v)$ to the graph.
The edge set $F$ is thus defined as $F := \{(e,v)~|~e \in E : \{v\} \subseteq e \}$.
When contracting a vertex pair $(u,v)$, we mark $v$ as deleted. The edges $(v,w)$ incident to $v$ are treated as follows:
If $G$ already contains an edge $(u,w)$, then net $w$ contained both $u$ and $v$ before the contraction. 
In this case, we simply delete the edge $(v,w)$ from the graph. Otherwise, net $w$ only contained $v$. We therefore have to
relink the edge $(v,w)$ to $u$.
Representing this graph using an adjacency array allows us to implement deletion and relink operations with very little space overhead.

After initial partitioning, we initialize the connectivity set $\mathrm{\Lambda}(e)$ as well as the pin counts $\mathrm{\Phi}(e,V_i)$ for each
cut net $e$. These data structures are then maintained and updated during the local search phase.

\subsection{Coarsening} \label{Coarsening}
The vertex pairs $(u,v)$ to be contracted are chosen according to a rating function. The goal of the coarsening phase
is to contract highly connected vertices such that the number of nets remaining in the hypergraph and their size is
successively reduced \cite{hMetisRB}. Removing nets leads to simpler instances for initial partitioning, while small net sizes allow
 FM-based local search algorithms to identify moves that improve the solution quality. Our coarsening algorithm therefore prefers vertex 
pairs that have a large number of heavy nets with small size in common. This score is then inversely scaled with
the product of the vertex weights $c(v)$ and $c(u)$ to keep the vertex weights of the coarse hypergraph reasonably uniform: 

\begin{equation}
r(u,v) := \frac{1}{{c(v) \cdot c(u)}}~\sum \limits_{e \in \{I(v) \cap I(u)\}}  \frac{\omega(e)}{|e| - 1}.
\end{equation}

This scaling factor was already effective in $n$-level graph partitioning~\cite{nGP}.
At the beginning of the coarsening algorithm, all vertices are rated in random order, i.e., for each vertex $u$ we 
compute the ratings of all neighbors $\mathrm{\Gamma}(u)$ and choose the vertex $v$ with the highest rating as contraction partner for $u$.
Ties are broken randomly. For each vertex, we insert the vertex pair with the highest score into an addressable PQ 
using the rating score as key. This allows us to efficiently choose the next vertex pair that should be contracted. After contraction, we remove 
$v$ from the PQ. We then remove all parallel- and single-node nets in $I(u)$. 
The latter are easily identified, because $|e|=1$. For parallel hyperedge detection we use an efficient algorithm similar to the one in \cite{ParallelHEDetection},
which is used to identify vertices with identical structure in a graph: 
We create a fingerprint for each net $e$ : $f_i := \bigoplus_{v \in e} v\oplus x$, for some seed $x$. These fingerprints are then sorted, which brings potentially
parallel nets together. A final scan over the fingerprints then identifies parallel nets: Only if two consecutive fingerprints $f_i, f_j$ are 
identical, we have to check whether $e_i~\triangle~e_j = \emptyset$ by comparing their pins.

Since each contraction potentially influences the rating scores of all neighbors $\mathrm{\Gamma}(u)$, we have to recalculate their ratings 
and update the priority queue accordingly. To avoid unbalanced inputs for the initial partitioning phase, vertices $v$ 
with $c(v) > c_{max} := s  \cdot \frac{c(V)}{t}$ are never allowed to participate in a contraction step and are thus removed from the priority queue.
The parameter $s$ will be chosen in \autoref{AlgorithmConfiguration} and $t$ is the maximum size of the coarsest graph, which we set to $160k$. 
We refer to this algorithm as \emph{full}.

As in the graph partitioning case, the $n$-level approach has the advantage that it obviates the need to employ a matching or clustering
algorithm to determine the vertices to be contracted. However, this comes at the expense of continuously re-rating the neighbors $\mathrm{\Gamma}(u)$ adjacent to the representative. In hypergraph partitioning, this is the most expensive part of the algorithm, because after each contraction, we have to look at all pins of all incident nets $I(v)$. The re-rating can therefore easily become the bottleneck -- especially if $H$ contains large nets. To improve the running time of the coarsening phase in these cases, we developed two variations of the full algorithm. 
Both variations only differ in the 
way the re-rating of adjacent vertices is handled. After contracting the vertex pair $(u,v)$, the first version only updates the rating of those 
neighbors, which had chosen either the representative $u$ or the contracted vertex $v$ as contraction partner. This can be done efficiently by 
maintaining the set $L_w := \{u~|~ (u,w) \in \text{PQ}\}$ of all representatives that choose $w$ as contraction partner. The re-rating step then only reevaluates the rating function 
for each vertex in $L_u \cup L_v$. All other ratings are left untouched. We refer to this version as \emph{partial}.
The second variation does not re-rate any vertices immediately after the contraction. Instead, all adjacent vertices $\mathrm{\Gamma}(u)$ are marked as \emph{invalid}. If the priority queue returns an invalid vertex, we recalculate its rating and update the priority queue accordingly. In case the queue returns a valid rating, we normally continue with the coarsening process.
This version is referred to as \emph{lazy}.

\subsection{Initial Partitioning} \label{InitialPartitioning}
The coarsening process is repeated until the number of remaining vertices is below $160k$ or the priority queue becomes empty. The latter can happen if no valid contraction step remains, e.g., a step that would not lead to a representative $u$ having weight $c(u) > c_{max}$. The hypergraph is then small enough to
be initially partitioned by an initial partitioning algorithm. Our framework allows using hMetis or PaToH as initial partitioner.
Because hMetis produces better initial partitions than PaToH, we use the recursive bisection variant of hMetis for initial partitioning.
In this variant of hMetis, the maximum allowed imbalance of a partition is defined differently \cite{hMetisRB}: An imbalance value of 5, for example,
allows each block to weigh between $0.45 \cdot c(V)$ and $0.55 \cdot c(V)$ \emph{at each bisection step}. We therefore translate our maximum allowed block 
weight to match this definition, i.e., we use imbalance parameter 
\begin{equation} \label{eq:RBimbalance}
\varepsilon' := 100 \cdot \left(\left( \frac{1+\varepsilon}{k} + \frac{\max_{v \in V} c(v)}{c(V)}\right)^{\frac{1}{\log_2(k)}} - 0.5 \right)
\end{equation}
for initial partitioning with hMetis. We call the initial partitioner multiple times with different random seeds and use  the best partition as initial partition of the coarsest graph.

\subsection{Localized direct $k$-way FM Local Search} \label{localizedFM}
Our local search algorithm follows ideas similar to the $k$-way FM-algorithm proposed by Sanchis~\cite{HypergraphKFM} and is further inspired by the local search algorithm used by Sanders and Osipov~\cite{nGP}. Sanchis uses $k (k-1)$ priority queues
to be able to maintain all possible moves for all vertices. We reduce the number of priority queues to $k$ -- one queue $P_i$
for each block $V_i$. In contrast to Sanchis, we only consider to move a vertex to \emph{adjacent} blocks rather than calculating
and maintaining gains for moves to \emph{all} blocks. This simultaneously reduces the memory requirements and restricts the search
space of the algorithm to moves that are more likely to improve the solution. Another key difference is the way a local search pass
is started: Instead of initializing the priority queues with all vertices or all border vertices, we perform a highly localized 
local search starting only with the represenative and the just uncontracted vertex. The search then gradually expands around
this vertex pair by successively inserting moves for neighboring vertices into the queues.

\paragraph{Algorithm Outline.} 
At the beginning of a local search pass, all
queues are empty and disabled. A disabled priority queue will not be considered when searching for the next move with the highest gain. All vertices are labeled inactive and unmarked. Only unmarked vertices are allowed to 
become active. To start the local search phase after each uncontraction, we activate the representative and the just uncontracted vertex if they are border vertices. Otherwise, no local search phase is started. 
\emph{Activating} a vertex $v$ means that we calculate the \emph{gain} $g_{i}(v)$ for moving $v$ to all adjacent blocks $i \in R(v) \setminus \{b[v]\}$ and 
insert  $v$ into the corresponding queues $P_i$ using $g_i(v)$ as key. The gain $g_i(v)$ is defined as: 
\begin{equation} \label{eq:gain}
g_{i}(v) := \sum \limits_{e \in I(v)} \{ \omega(e) : \mathrm{\Phi}(e, i) = |e| - 1\} - \sum \limits_{e \in I(v)} \{ \omega(e) : \lambda(e) = 1\}.
\end{equation}
Thus, instead of considering all $k - 1$ possible moves of a vertex $v$, we only examine moves to those blocks that are in the union of the connectivity sets of its incident nets: $\bigcup_{e \in I(v)} \{\mathrm{\Lambda}(e) \setminus b[v]\} $. After insertion, all PQs corresponding to \emph{underloaded} blocks become enabled. Since a move to an overloaded block will
never be feasible, all queues corresponding to overloaded blocks are left disabled. The algorithm then 
repeatedly queries only the \emph{non-empty, enabled} queues to find the move with the highest gain $g_{i}(v)$, breaking ties randomly. 
Vertex $v$ is then moved to block $V_i$ and labeled inactive and marked. Since each vertex is allowed to move at most once during each pass, we remove all other moves of $v$ from the PQs.
We then update all neighbors of $v$ and continue local search until either no non-empty, enabled PQ remains or a constant number of $c$ moves neither decreased the cut nor improved the current imbalance. The latter criterion is necessary, because otherwise the $n$-level 
approach could lead to $|V|^2$ local search steps in total. 
After local search is stopped, we undo all moves until we arrive at the lowest cut state reached during the search that fulfills the balance constraint. 
All vertices become unmarked and inactive and the algorithm is then repeated until no further improvement is achieved.

\paragraph{Activation and Gain Computation.}

\begin{algorithm2e}[h]
\caption{Activation}\label{alg:gain}\normalsize
\Function{\Activate{$v$,$G$}} {
\KwIn{Vertex $v$ to be activated, gain array $\mathrm{\Omega}[1..k]$, $\forall 1 \leq i \leq k : \mathrm{\Omega}[i] = 0$}
\lIf(\Remi{only activate border vertices}) {$v$ \upshape is not a border vertex} {\Return}
$\omega_{int} := 0$ \Rem{weight of internal nets in $I(v)$}
$R := \{\}$ \Rem{discovered blocks adjacent to $v$}
\ForEach(\Remi{visit incident nets}) {$e \in I(v)$} {
  \Switch(\Remi{and look at connectivity}){$\mathrm{\lambda}(e)$}{
    \Case(\Remi{$e$ is internal}){$1:$}{
      \lIf() {$|e| > 1$} {  
      $\omega_{int} :=  \omega_{int} + \omega(e)$
      }
    }
    \Case(\Remi{$e$ might be removable from the cut}){$2:$}{
      \ForEach(\Remi{visit all connected blocks}) {$ V_i \in \mathrm{\Lambda}(e)$} {
        $R := R \cup \{V_i\}$ \nllabel{alg:nHGP:activate:r1}

        \If(\Remi{move removes $e$ from the cut.})
        {$\mathrm{\Phi}(e,V_i) = |e| - 1$}
            {
              $\mathrm{\Omega}[V_i] := \mathrm{\Omega}[V_i] + \omega(e)$
            }
        }
    }
    \Other(\Remi{$e$ will not be removable from the cut}){
      \ForEach() {$ V_i \in \mathrm{\Lambda}(e)$} {
        $R := R \cup \{V_i\}$ \Remi{to find moves with negative or zero gain \nllabel{alg:nHGP:activate:r2}}
      } 
    } 
  }
}
$R := R \setminus \{b[v]\}$ \Rem{remove current block of $v$}
$\mathrm{\Omega}[b[v]] := 0$ \Remi{and reset the gain value}

\ForEach(\Remi{$R$ now contains all adjacent blocks}) {$V_i \in R$} {
  $P_i$.insert($v$, $\mathrm{\Omega}[V_i] - \omega_{int}$) \Remi{$g_{i}(v) = \mathrm{\Omega}[V_i] - \omega_{int}$}\nllabel{alg:nHGP:activate:gain}

  $\mathrm{\Omega}[V_i] := 0$ \Remi{reset slot to initial state}

  \lIf(\Remi{enable eligible PQs}) {$c(V_i) < L_{\text{max}}$} {
    $P_i$.enable()
    }
}
\KwOut{$\forall $ adjacent blocks $V_i$ of $v: P_i$ contains $v$ with priority $g_i(v)$. If block $V_i$ is underloaded, priority queue $P_i$ is enabled.}
}
\end{algorithm2e}

Our gain computation algorithm is detailed in Algorithm~\ref{alg:gain}. For each vertex $v$, we are only interested in gain values for moves to adjacent blocks $R(v)$. Calculating these gains can be done efficiently by looking at all incident nets $I(v)$ and all adjacent blocks \emph{exactly once}. While iterating over $I(v)$, we generate the set $R(v)$ of all adjacent blocks (lines~\ref{alg:nHGP:activate:r1} and \ref{alg:nHGP:activate:r2}). To calculate the gain, we distinguish three cases for each net $e$: If $\mathrm{\lambda}(e)=1$ and $|e| > 1$,
net $e$ is internal in block $b[v]$ and will become a cut net when $v$ is moved to another block. We maintain the sum of the weights of all internal nets in $\omega_{int}$. If $\mathrm{\lambda}(e)=2$ and one block $V_i$ of the two blocks in the connectivity set $\mathrm{\Lambda}(e)$ contains all but one pin, net $e$ can be removed from the cut by moving $v$ to block $V_i$.
The weight of these nets is stored in  $\mathrm{\Omega}[V_i]$. All other nets cannot be
removed from the cut by moving $v$ to a different block. We therefore just update $R$ accordingly.
Finally, by iterating over the set of all adjacent blocks, we can compute the gain values for moving $v$ to all connected blocks $V_i$ by substracting the internal weight $\omega_{int}$ from the weight stored in $\mathrm{\Omega}[V_i]$ (line~\ref{alg:nHGP:activate:gain}).

\paragraph{Update of Neighbors.} \label{par:gainupdate}
\begin{algorithm2e}[h]
\caption{Delta-Gain-Update}\label{alg:gainupdate}\normalsize
\Function{\DeltaGainUpdate{$v,V_{\text{\upshape from}},V_{\text{\upshape to}}$}} {
\KwIn{Vertex $v$ that was moved from block $V_{\text{\upshape from}}$ to $V_{\text{\upshape to}}$}
\ForEach(\Remi{walk all incident nets}) {$e \in I(v)$} {
  \ForEach(\Remi{and consider each pin}\nllabel{alg:gainupdate:forallpins}) {$ u \in e \setminus \{v\}$} {
    \If(\Remi{move made $e$ a cut net}\nllabel{alg:gainupdate:case1beg}) {$\mathrm{\Phi}(e,V_{\text{\upshape from}}) = |e| - 1 $} {

      \lForEach() {$V_i \in \mathrm{\Pi} \setminus \{V_{\text{\upshape from}}\}$} {
        $P_{i}.\FuncSty{update}(u, \omega(e))$ \nllabel{alg:gainupdate:case1end}
      }
    }
    
    \If(\Remi{move removed $e$ from the cut}\nllabel{alg:gainupdate:case2beg}) {$\mathrm{\Phi}(e,V_{\text{\upshape to}})  = |e|$} {

      \lForEach() {$V_i \in \mathrm{\Pi} \setminus \{V_{\text{\upshape to}}\}$} {
        $P_{i}.\FuncSty{update}(u, -\omega(e))$\nllabel{alg:gainupdate:case2end}
      }
    }
    
    \If(\Remi{only $v$ still outside $V_{\text{\upshape to}}$}\nllabel{alg:gainupdate:case3beg}) {$\mathrm{\Phi}(e,V_{\text{\upshape to}}) = |e| -1 \wedge b[u] \neq V_{\text{\upshape to}} $} {      
      \emph{// moving it to $V_{\text{\upshape to}}$ would remove $e$ from the cut}

      $P_{\text{\upshape to}}.\FuncSty{update}(u, \omega(e))$ \nllabel{alg:gainupdate:case3end}
    }
    
    \If(\Remi{$2$ pins outside $V_{\text{\upshape from}}$}\nllabel{alg:gainupdate:case4beg}) {$\mathrm{\Phi}(e,V_{\text{\upshape from}}) = |e| - 2 \wedge b[u] \neq V_{\text{\upshape from}} $} {
      \emph{// moving $v$ to $V_{\text{\upshape from}}$ could have removed $e$ from the cut}

      $P_{\text{\upshape from}}.\FuncSty{update}(u, -\omega(e))$ \nllabel{alg:gainupdate:case4end}
    }
  }
}
\KwOut{The gains for all moves of all neighbors $\mathrm{\Gamma}(v)$ are updated.}
}
\end{algorithm2e}

After moving a vertex $v$ from block $V_{\text{from}}$ to a different block $V_{\text{to}}$, we have to update all of its neighbors  $\mathrm{\Gamma}(v)$. 
All previously inactive neighbors are activated using Algorithm~\ref{alg:gain}. 
All neighbors that became internal are labeled inactive and all corresponding moves are deleted from the priority queue.
Finally,  we update the gains for all moves of the remaining neighbors that are already active and remain border vertices. 
We reuse the gain values that are already calculated and only perform \emph{delta-gain-updates}: 
If the move changed the contribution to the gain values for a net $e \in I(v)$, we
account for that change by incrementing/decrementing the gains of the corresponding moves by $\omega(e)$. 
Our delta-gain-update algorithm is outlined in Algorithm~\ref{alg:gainupdate}. For each net $e$, we have to consider four cases: \sschl{For TR, i could make a picture to show these cases}

\begin{enumerate}
\item Before the move, net $e$ was completely internal in block $V_{\text{from}}$. Now, after the move, $e$ has become a cut net with $\lambda(e) = 2$ and $\mathrm{\Lambda}(e) = \{V_{\text{from}}, V_{\text{to}}\}$, if all but one pin of net $e$ are in block $V_{\text{from}}$.
  In this case, before the move, net $e$ contributed $-\omega(e)$ to the gain of all its pins for moving to another block.
  Since the move of vertex $v$ now made $e$ a cut net, all other pins can be moved to another block without incurring a further increase in cut. We therefore change the contribution
  of net $e$ from $-\omega(e)$ to zero by increasing the corresponding gains by $\omega(e)$ (lines~\ref{alg:gainupdate:case1beg}~to~\ref{alg:gainupdate:case1end}).
\item Vertex $v$ was the only pin of net $e$ that was outside of block $V_{\text{to}}$ before the move and the movement therefore removed $e$ from the cut. 
  Now that $e$ is internal, it contributes $-\omega(e)$ to the gain of all its remaining pins for moving to another part, 
  since each move would again make it a cut net (lines~\ref{alg:gainupdate:case2beg}~to~\ref{alg:gainupdate:case2end}).
\item After the move of $v$ only one pin of net $e$ remains outside of $V_{\text{to}}$. If that pin is also moved to $V_{\text{to}}$, we remove $e$ from the cut. 
  The corresponding move of this pin therefore receives a delta-gain-update of $\omega(e)$. For all other pins, the contribution of $e$ to their gain values did not change 
  (lines~\ref{alg:gainupdate:case3beg}~to~\ref{alg:gainupdate:case3end}).
\item Before the move of $v$, there was only one pin left that was outside of $V_{\text{from}}$. Moving this pin to $V_{\text{from}}$ would have removed $e$ from the cut.
  However, now that $v$ is also outside of $V_{\text{from}}$, the move of this pin cannot decrease the cut any more. The contribution of net $e$ to the gain of moving this pin to $V_{\text{from}}$ 
  therefore changes from $\omega(e)$ to zero (lines~\ref{alg:gainupdate:case4beg}~to~\ref{alg:gainupdate:case4end}).
\end{enumerate}

For each active vertex, the priority queues maintain the gains to all \emph{adjacent} blocks.
The set of adjacent blocks, however, is subject to change during local search, because vertex movements can increase as
well as decrease the connectivity of incident nets. The update process therefore has to take these changes into account.
Otherwise we would either miss potential moves or perform \emph{stale} moves, i.e., move a vertex to a block
that was adjacent to $v$ at some point of the local search but is not adjacent any more at the time it is returned by the priority queue.
The movement of $v$ increased the set of adjacent blocks $\mathrm{R}(u)$ for one of its neighbors $u$ if $V_{\text{to}} \not\in \mathrm{R}(u)$ before the move.
In this case, we calculate the gain $g_{\text{to}}(u)$ and insert $u$ into the priority queue $P_{\text{to}}$. Similarly, if the movement decreased the set of
adjacent blocks (i.e., $V_{\text{from}} \not\in \mathrm{R}(u)$ after the move), we remove the vertex from $P_{\text{from}}$.

\paragraph{Critical Nets.}
Having identified the cases in which a move changes the gain contribution of one of its incident nets,
we generalize the notion of \emph{critical nets} introduced by Fiduccia and Mattheyses~\cite{FM82} for bipartitioning to $k$-way partitioning. 
A net is said to be \emph{critical} if there exists at least one move for one of its pins that affects the gain contribution. 
Notice that the conditions concerning $\mathrm{\Phi}(e,\cdot)$ in lines~\ref{alg:gainupdate:case1beg}, \ref{alg:gainupdate:case2beg}, \ref{alg:gainupdate:case3beg} and \ref{alg:gainupdate:case4beg} of
Algorithm~\ref{alg:gainupdate} can be evaluated without considering a pin of the corresponding net. Thus, we can determine whether
or not a net is critical by evaluating these conditions once before iterating over all pins (line~\ref{alg:gainupdate:forallpins}). Only if one of the conditions evaluates to true, we actually have to consider each pin of the net.     

\paragraph{Locked Nets.}
Performing delta-gain-updates rather than re-calculating the gains for each neighbor of a moved vertex from scratch considerably reduces the complexity of the update step.
The complexity can be reduced even further by noticing that the contribution of a net does not change any more once two of its pins have been
moved to two \emph{different} blocks. The net is then \emph{locked} in those two blocks, because neither of the two vertices is allowed to be
moved again during the current local search pass. It is therefore not possible to remove such a net from the cut by moving any of the remaining movable pins to another block. Thus it is not necessary to perform any further delta-gain-updates for locked nets.
This observation was first described by Krishnamurthy~\cite{LockedNets} for bipartitioning and transferred to $k$-way partitioning by Sanchis~\cite{HypergraphKFM}.
We integrate locking of nets into our algorithm by labeling each net during a local search pass. Initially, all nets are labeled \emph{free}.
Once the first pin of a net is moved, the net becomes \emph{loose}. It now has a pin in one block that cannot be moved again. Further moves to this 
block do not change the label of the net. As soon as another pin is moved to a different block, the net is labeled \emph{locked} and is
excluded from future delta-gain-updates. However, we still have to account for changes in connectivity as described above.

\subsection{Local Search with Size-Constrained Label Propagation (SCLaP)} \label{LPAlocalSearch}
The \emph{label propagation algorithm} for graph
clustering was recently equipped with a size constraint to work as a coarsening and a local search
algorithm for graph partitioning~\cite{LPAgraphPartitioning}. 
We briefly outline the previous local search algorithm before describing our adaptation to hypergraphs.
Initially, each vertex $v$ is in its current block $b[v]$.
The algorithm then works in rounds. In each round, the vertices are visited in random order and each
vertex $v$ is moved to the eligible (\ie not overloaded after the move) block $V_i$ 
that has the strongest connection to $v$.
Ties are broken randomly. 
After all vertices are visited, the process is repeated until the labels have converged or
a maximum number of $\ell$ rounds is reached. 

\begin{algorithm2e}[!b]
\caption{Label Propagation Local Search}\label{alg:lparefine}\normalsize
\KwIn{Uncontracted vertex pair $(u,v)$}
$Q_1 := \{u, v \}$ \Remi{set of vertices for current iteration}

$Q_2 := \{ \}$ \Remi{set of vertices for next iteration}

\While {$Q_1 \neq \{ \} \wedge num\_iterations \leq max\_iterations$} {
   \ForEach {$v \in Q_1 $ \upshape in random order}{
     $\mathrm{\Omega}[1..k] := \bot$

     \ForEach(\Remi{for all adjacent blocks}\nllabel{alg:gainLPA:beg}) {$V_i \in R(v) $}{
       \ForEach(\Remi{calculate gains for moves}) {$e \in I(v) $}{
         \If(\Remi{enforce balance constraint}) {$c(v) + c(V_i) \leq L_{\text{max}}$}{
           \lIf {$\mathrm{\Omega}[V_i] = \bot$} { $\mathrm{\Omega}[V_i] := 0$} 
           $\mathrm{\Omega}[V_i] := \mathrm{\Omega}[V_i] + \FuncSty{gain}(v,e,V_i)$\nllabel{alg:gainLPA:end}
         }
       }
     }
     $V_{max} := \argmax_{V_i}{\mathrm{\Omega}[V_i]}$ \Remi{choose max-gain move with max. $\mathrm{\Lambda}$-decrease}
     
     \If{$b[v] \neq V_{max}$} {
       $\FuncSty{move}(v,V_{max})$

       \ForEach()  {$w \in \mathrm{\Gamma}(v)$} {
         \emph{// all neighbors may change their block in next round}
         $Q_2 := Q_2 \cup \{w\}$
       }
     }
   }
   $Q_1 := \{\}$
   
   $\FuncSty{swap}(Q_1,Q_2)$
 } 

\KwOut{Refined partition  $\mathrm{\Pi} = \{V_1, \dots, V_k\}$}
\end{algorithm2e}

We modify this local search algorithm as follows in order to be applicable in our $n$-level hypergraph partitioning context (see Algorithm~\ref{alg:lparefine}).
Instead of iterating over all vertices, we start the first iteration only with the vertex pair $(u,v)$ that has just been
uncontracted. If one of these two vertices changes its block, all of its neighbors
are allowed to change their block in the next iteration. This can be done efficiently by maintaining  two
queues $Q_1$ and $Q_2$. $Q_1$ contains the vertices for the current iteration and $Q_2$ those 
for the next iteration. After each round, we clear $Q_1$ and swap it with $Q_2$.
In order to reflect our partitioning objective, we move a vertex to the eligible block that maximizes the
\emph{gain} as defined in \autoref{eq:gain}. For each incident net $e$, we calculate its contribution to the gain
for moving it to each adjacent block $V_i$ (\autoref{alg:gainLPA:beg}~--~\autoref{alg:gainLPA:end}): 

\begin{equation}
\FuncSty{gain}(v,e,V_i) :=  \left\{ 
  \begin{array}{r l}
    -\omega(e) & \quad \text{if $\mathrm{\lambda}(e)= 1 \wedge V_i \not\in \mathrm{\Lambda}(e)$}\\
    \omega(e)  & \quad \text{if $\mathrm{\lambda}(e)= 2 \wedge \mathrm{\Phi}(e,V_i)=|e| - 1 $} \\
    0          & \quad \text{else}
  \end{array} \right.
\end{equation}

Finally, we adapt the tie-breaking scheme: If multiple blocks have the same maximum gain, we choose
to move the vertex to the block that leads to the highest connectivity decrease for all incident nets.
A move of vertex $v$ to block $V_i$ decreases the connectivity of a net $e$, if $\mathrm{\Phi}(e, b[v]) = 1$ 
and $\mathrm{\Phi}(e, V_i) \neq 0$. Analogously, a move increases the connectivity if $\mathrm{\Phi}(e, V_i) = 0$.
The total connectivity decrease for moving a vertex $v$ to block $V_i$ can therefore be computed during the gain calculation. 
The intention behind this tie-breaking scheme is to
successively reduce the number of blocks a net is connected to. Only in case multiple blocks also have the same 
connectivity decrease value, we resort to random tie breaking.

\subsection{Iterated Multilevel Algorithms}  \label{GlobalSearch}
\emph{V-cycles} are a common technique 
to further improve a solution~\cite{hMetisRB,KahipWFCycles,WalshawVcycle}. The idea is to reuse an already computed partition as input for the multilevel approach. 
During coarsening, the quality of the solution is maintained by only contracting vertices
belonging to the same block. The current partition of the coarsest graph is then used as initial partition . During uncoarsening, local search 
algorithms can then further improve solution quality. 
We also adopt this technique for $n$-level hypergraph partitioning by modifying the rating algorithm 
such that we only allow the contraction of vertex pairs that belong to the same block. 

\section{Experiments} \label{Experiments}
\paragraph{Instances.} \label{Instances}
We evaluate our algorithms on hypergraphs derived from two well established benchmark sets: The ISPD98 Circuit Benchmark Suite~\cite{ISPD98} and
the University of Florida Sparse Matrix Collection~\cite{FloridaSPM}. From the latter, we use the instances that are part of the 10th DIMACS implementation challenge dataset~\cite{benchmarksfornetworksanalysis}. The matrices are translated into hypergraphs using the row-net model, i.e. each row of the matrix
is treated as a net. All hypergraph have unit net and vertex weights. We exclude the two largest instances \emph{nlpkkt200} and \emph{nlpkkt240}, because they could not 
be partitioned using hMetis. In both cases the amount of memory needed by hMetis exceeded the amount of memory available on our machine.

We divided the benchmark set into \emph{medium-sized} and \emph{large} instances and use $k \in \{2,4,8,16,32,64,128\}$ for the number of blocks and an allowed imbalance of $\varepsilon = 0.03$.
The properties of the hypergraphs are summarized in \autoref{tbl:mediuminstances}. 

\paragraph{System.}
All experiments are performed on a single core of a machine consisting of two Intel Xeon E5-2670 Octa-Core processors (Sandy Bridge) 
clocked at $2.6$ GHz. The machine has $64$ GB main memory, $20$ MB L3-Cache and 8x256 KB L2-Cache and is running Ret Hat Enterprise Linux (RHEL) 6.4.

\paragraph{Methodology.}  \label{Methodology}
The algorithms are implemented in the $n$-level hypergraph partitioning framework \emph{KaHyPar} (\textbf{Ka}rlsruhe \textbf{Hy}pergraph \textbf{Par}titioning). The code is written in C++ and compiled using gcc-4.9.1 with flags \texttt{-O3} \texttt{-mtune=native} \texttt{-march=native}. 
Unless otherwise mentioned, we perform ten repetitions with different seeds for each experiment and report the arithmetic 
mean of the computed cut and running time as well as the best cut found. When averaging over different instances, we use the 
geometric mean in order to give every instance a comparable influence on the final result. 
We compare our algorithms to both the $k$-way (hMetis-K) and the recursive-bisection variant (hMetis-R) of hMetis 2.0 (p1)~\cite{hMetisRB,hMetisKway} and to PaToH 3.2~\cite{PaToH}. 
As noted in \autoref{InitialPartitioning}, hMetis-R employs a different balance constraint. We therefore translate our
imbalance parameter $\varepsilon=0.03$ to $\varepsilon'$ as described in \autoref{eq:RBimbalance} such that it matches our balance constraint after $\log_2(k)$ 
bisections.  
Since PaToH ignores the random seed if configured to use the quality preset, we report both the result of the quality preset (PaToH-Q) as well as the
average over ten repetitions using PaToH in default configuration (PaToH-D). In both cases, we configured PaToH to use a final imbalance ratio of $\varepsilon = 0.03$ to
match our balance constraint.

\paragraph{Algorithm Configuration.}  \label{AlgorithmConfiguration}
We performed a large amount of experiments to tune the parameters of our algorithms using the medium-sized instances and $k \in \{2,4,8,16,32,64\}$. 
A full description of these experiments is omitted due to space constraints. We use two sets of parameter settings \emph{fast} and \emph{strong} that turned out to work well. Both configurations use the same parameters for coarsening and initial partitioning:  We use the \emph{lazy} variant 
as coarsening algorithm, because it is the fastest algorithm and provides comparable quality to both other variants (see \autoref{tbl:coarsening}).
The coarsening process is stopped as soon as the number of vertices drops below $t=160k$ or no eligible vertex is left. 
The scaling factor $s$ for the maximum allowed vertex weight during coarsening is set to $2.5$. We use the default configuration of hMetis as initial partitioner and perform only one initial partitioning trial, since more iterations seldomly produced different cuts.
The strong configuration uses the $k$-way FM algorithm described in \autoref{localizedFM} and stops each pass as soon as $c=200$ moves did not yield any improvement.
The fast configuration employs the SCLaP-based refinement algorithm presented in \autoref{LPAlocalSearch} and performs $\ell=5$ rounds on each level. 
Both configurations can be augmented using V-cycles (referred to as \emph{FastV} and \emph{StrongV}). 
The maximum number of V-cycle iterations is set to $3$ for FastV and to $10$ for StrongV.

\begin{SCtable}
\centering
\caption{Test results for the three different variants of our coarsening algorithm on medium-sized instances. Running time of the coarsening phase is relative to \emph{full}.}
\label{tbl:coarsening}
\begin{tabular}{lccccc}
Variant     & avg. cut &  & best cut  & & $\frac{\text{coarsening time}}{\text{coarsening time}_{full}}$ \\
\hline
\hline
full        & \numprint{2990.95}  &    & \numprint{2910.30} &    & \numprint{1.00}             \\
partial   & \numprint{2996.93}  &   & \numprint{2914.21}  &   & \numprint{0.10}          \\
lazy        & \numprint{2988.74}  &    & \numprint{2916.62} &    & \numprint{0.06}          \\
\end{tabular}
\end{SCtable}

\subsection{Main Results}
\paragraph{Evaluation of Local Search Algorithms.}
In the following, we report the final results of the parameter tuning on the medium-sized benchmark set and 
evaluate the performance of our local search algorithms described in \autoref{localizedFM} and \autoref{LPAlocalSearch} in the $n$-level context.
The results are summarized in \autoref{tbl:FastvsStrongDetailed}. Using an FM-based heuristic pays off: 
The localized $k$-way FM algorithm consistently outperforms the greedy SCLaP-based algorithm. The cuts of label propagation using V-cycles
are on average $4\%$ larger than those of StrongV.
Its advantage decreases for larger values of $k$. This can be explained by the fact that as $k$ increases, cut nets are more likely to connect
more than two blocks. The FM algorithm then has to deal with an increasing number of zero-gain moves, which effectively weakens its
ability to climb out of local minima. \csch{mention somewhere that small improvments in the objective have a large value in practice with the right citation (perhaps in the intro)}

V-cycles improve the solution quality of both algorithms by around $1\%$ on average. The impact of global search iterations 
is larger for the SCLaP-based algorithm than for localized $k$-way local search, especially if the number of blocks is small.
In these cases, it is more likely that a vertex can switch its label and thereby improve the solution quality. With increasing $k$,
this becomes more difficult. The effect of V-Cycles is more stable for $k$-way FM algorithm, since it is already a strong heuristic.   

Considering running times, we note that the SCLaP-based algorithm is an order of magnitude faster on average than the localized FM algorithm.
However, best cuts found by the former are still larger than the average cuts of the latter -- even for instances of medium size.

\begin{table}[h]
\centering
\caption{Performance of our local search algorithms on the medium-sized instances used for parameter tuning.
All average cuts and best cuts are shown as increases (\%) relative to the values obtained by StrongV.}
\label{tbl:FastvsStrongDetailed}
\begin{tabular}{rc|rrcr|rrcr|rrcr|rrrr}
\multirow{2}{*}{k} & & \multicolumn{4}{c|}{StrongV}  &\multicolumn{4}{c|}{Strong}   &\multicolumn{4}{c|}{FastV}    &\multicolumn{4}{c}{Fast}\\
    && \multicolumn{1}{c}{best}     & \multicolumn{1}{c}{avg}  & &\multicolumn{1}{c|}{t[s]}   &\multicolumn{1}{c}{best[\%]}  & \multicolumn{1}{c}{avg[\%]} &&\multicolumn{1}{c|}{t[s]}   &\multicolumn{1}{c}{best[\%]} & \multicolumn{1}{c}{avg[\%]} & &\multicolumn{1}{c|}{t[s]}   &\multicolumn{1}{c}{best[\%]} & \multicolumn{1}{c}{avg[\%]} & &\multicolumn{1}{c}{t[s]} \\
\hline
2   &&\numprint{792}	& \numprint{815}  & &\numprint{13.7}    &+0.56      &+0.52	&& 5.6	 & +4.50 & +5.70& & 1.3  &   +6.15 & +8.39 && 0.7  \\
4   &&\numprint{1662}	& \numprint{1744} & &\numprint{37.4}    &+1.54      &+1.02	&& 11.6	 & +5.17 & +5.31& & 1.9  &   +7.95 & +7.77 && 1.1  \\
8   &&\numprint{2823}	& \numprint{2920} & &\numprint{76.3}    &+1.66      &+1.18	&& 20.3	 & +5.43 & +5.52& & 2.8  &   +7.58 & +7.68 && 2.0  \\
16  &&\numprint{4090}	& \numprint{4190} & &\numprint{155.7}   &+1.54      &+1.45	&& 33.5	 & +4.57 & +4.50& & 4.6  &   +5.95 & +5.94 && 3.7  \\
32  &&\numprint{5454}	& \numprint{5529} & &\numprint{229.1}   &+1.17      &+1.06	&& 45.3	 & +3.32 & +3.18& & 7.8  &   +4.19 & +3.98 && 7.0  \\
64  &&\numprint{6843}	& \numprint{6921} & &\numprint{228.8}   &+0.92      &+0.85	&& 51.2	 & +2.42 & +2.32& & 14.1 &   +2.84 & +2.76 && 13.3 \\
\hline
avg && \numprint{2877}  &\numprint{2955}  & &82.7   &+1.23      &+1.01   && 21.6 & +4.23 & +4.41& & 3.9     &   +5.76 & +6.07 && 2.8 \\
\end{tabular}
\end{table}

\paragraph{Comparison to other Hypergraph Partitioners.}
\begin{table}[h]
\centering
\caption{Comparison of our algorithms and state-of-the-art hypergraph partitioners on large benchmark instances.}
\label{tbl:comparisonOverview}
\begin{tabular}{lcrcccrcccrS[table-format=3.2]}
Algorithm && \multicolumn{1}{c}{avg. cut} &&&& \multicolumn{1}{c}{best cut}  &&&& \multicolumn{1}{c}{t[s]} \\
\hline
StrongV	  &&	\numprint{12968}	&&&&	\numprint{12706}	&&&&	979.0	\\
Strong	  &&	\numprint{13088}	&&&&	\numprint{12815}	&&&&	211.3	\\
FastV	  &&	\numprint{13955}	&&&&	\numprint{13581}	&&&&	61.5	\\
Fast	  &&	\numprint{14269}	&&&&	\numprint{13861}	&&&&	30.7	\\
\hline
hMetis-R  &&	\numprint{13155}	&&&&	\numprint{12977}	&&&&	230.7	\\
hMetis-K  &&	\numprint{13548}	&&&&	\numprint{13341}	&&&&	134.3	\\
PaToH-Q	  &&	\numprint{13805}	&&&&	\numprint{13805}	&&&&	12.6	\\
PaToH-D	  &&	\numprint{14560}	&&&&	\numprint{13912}	&&&&	3.1	\\
\end{tabular}
\end{table}

We now switch to our benchmark set containing large instances to avoid the effect of overtuning our algorithms to
the instances used for parameter tuning.  We exclude \emph{cage15} from the following results, because hMetis took more than 18 hours to compute a single bipartition.
hMetis-K is the only algorithm that often produces imbalanced partitions. Out of 1610 cases, 209 partitions are imbalanced (up to $12\%$ imbalance).
It therefore has slight advantages in the following comparisons because we do not disqualify
imbalanced partitions.
 \autoref{tbl:comparisonOverview} and \autoref{tbl:comparisonStrongByK} summarize the results. Detailed per-instance results can be found in the Appendix in \autoref{tbl:detailedResults}.

The strong variants of KaHyPar produce the smallest average and minimum cuts. The results of KaHyPar-FastV are comparable to the cuts produced by PaToH.
On average, the cuts produces by PaToH-D, PaToH-Q, hMetis-K, hMetis-R are $12\%$, $7\%$, $5\%$ and $2\%$ larger than those of KaHyPar-StrongV, respectively.
hMetis-R performs surprisingly well, considering the fact that we had to tighten the balancing constraint in order to ensure 
balanced solutions. However, out of 161 instances, KaHyPar-StrongV computed 112 partitions that were better than those
produced by hMetis-R and reproduced the cuts of hMetis-R in 14 of the remaining 49 cases. 
Also note that KaHyPar-Strong dominates the previously best solver hMetis-R with respect to both quality and running time.

As can be seen in \autoref{tbl:comparisonStrongByK}, the greedy label propagation algorithm outperforms PaToH on VLSI instances, while still being on average
around four times faster than hMetis. On sparse matrix instances however, it is not able to escape from local minima and thus cannot improve the quality
above PaToH's level.  

Looking at the more detailed comparison of KaHyPar-StrongV to the other partitioning packages in \autoref{tbl:comparisonStrongByK}, we see that the localized $k$-way
FM algorithm is only beaten by hMetis-R for $k=2$ and $k=4$. The improvement in solution quality increases with an increasing number of blocks. For $k=128$ our algorithm
produces $7\%$ better cuts than hMetis-K and $3\%$ better cuts than hMetis-R.
For large values of $k$, it becomes increasingly difficult for the greedy refinement algorithm used in hMetis-K to find moves with a positive gain.
The same problem holds true for our SCLaP-based algorithm, which actually performs worse than its counterpart in hMetis-K.
This could be explained by the fact that our algorithm only tries to optimize around the just uncontracted vertex pair and is likely to be trapped in a local
minimum, while the greedy refinement of hMetis-K in each iteration visits all vertices and moves them to an eligible block if the move has a positive gain. 

\begin{table}[!h]
\centering
\caption{Detailed comparison of our algorithms and other partitioners on large benchmark instances.
The first table summarizes the performance of the $k$-way local search algorithm on \emph{all} large instances.
The second and third table show the results for the greedy algorithm on large VLSI (second) and large sparse matrix instances (third).
The average cuts are shown as increases in cut (\%) relative to the values obtained by our algorithm shown in the first column.}
\label{tbl:comparisonStrongByK}
\begin{tabular}{rc|rcrc|rrc|rrc|rrc|rr}
\multirow{2}{*}{k} && \multicolumn{3}{c}{\textbf{StrongV}}                 & & \multicolumn{2}{c}{hMetis-K} & &  \multicolumn{2}{c}{hMetis-R} & & \multicolumn{2}{c}{PaToH-Q} &  &\multicolumn{2}{c}{PaToH-D} \\
    && avg. cut  &&  t[s] & &\multicolumn{1}{c}{cut [$\%$]} & \multicolumn{1}{c}{ t[s]} & & \multicolumn{1}{c}{cut [$\%$]} & \multicolumn{1}{c}{ t[s]} & &
\multicolumn{1}{c}{cut [$\%$]} & \multicolumn{1}{c}{ t[s]} & & \multicolumn{1}{c}{cut [$\%$]} & \multicolumn{1}{c}{ t[s]} \\
\hline
2  && \numprint{2563.6}  && \numprint{141.0}  && \numprint{+2.02} & \numprint{76.0}  && \numprint{-1.84} & \numprint{ 82.0} && \numprint{+2.94} & \numprint{4.0} && \numprint{+8.97} &	\numprint{1.1}	\\
4  && \numprint{5471.3}  && \numprint{313.9}  && \numprint{+3.02} & \numprint{89.7}  && \numprint{-0.07} & \numprint{151.4} && \numprint{+8.06} & \numprint{7.7} && \numprint{+13.60}&	\numprint{2.0}	\\
8  && \numprint{9382.7}  && \numprint{684.0}  && \numprint{+3.19} & \numprint{103.6} && \numprint{+1.22} & \numprint{211.7} && \numprint{+7.08} & \numprint{11.3}&& \numprint{+14.01}&	\numprint{2.8}	\\
16 && \numprint{14760.4} && \numprint{1100.0} && \numprint{+4.56} & \numprint{124.7} && \numprint{+2.27} & \numprint{267.5} && \numprint{+6.98} & \numprint{14.7}&& \numprint{+13.46}&	\numprint{3.5}	\\
32 && \numprint{21980.8} && \numprint{1884.3} && \numprint{+5.25} & \numprint{158.5} && \numprint{+2.44} & \numprint{319.9} && \numprint{+6.85} & \numprint{18.4}&& \numprint{+12.01}&	\numprint{4.3}	\\
64 && \numprint{32190.5} && \numprint{3115.7} && \numprint{+6.36} & \numprint{208.7} && \numprint{+2.75} & \numprint{369.6} && \numprint{+6.71} & \numprint{21.4}&& \numprint{+12.50}&	\numprint{5.0}	\\
128&& \numprint{44865.1} && \numprint{4407.8} && \numprint{+7.04} & \numprint{271.1} && \numprint{+3.46} & \numprint{418.8} && \numprint{+6.63} & \numprint{25.1}&& \numprint{+11.46}&	\numprint{5.7}	\\
\hline																					
avg&& \numprint{12967.7}  && \numprint{979.0}  && \numprint{+4.48} & \numprint{134.3} && \numprint{+1.45} & \numprint{230.7} && \numprint{+6.45} &\numprint{12.6} && \numprint{+12.28}&	\numprint{3.1}	\\
\end{tabular}

\vspace{1cm}

\begin{tabular}{rc|rcrc|rrc|rrc|rrc|rr}
\multirow{2}{*}{k} && \multicolumn{3}{c}{\textbf{FastV}}                 & & \multicolumn{2}{c}{hMetis-K} & &  \multicolumn{2}{c}{hMetis-R} & & \multicolumn{2}{c}{PaToH-Q} &  &\multicolumn{2}{c}{PaToH-D} \\
    && avg. cut  &&  t[s] & &\multicolumn{1}{c}{cut [$\%$]} & \multicolumn{1}{c}{ t[s]} & & \multicolumn{1}{c}{cut [$\%$]} & \multicolumn{1}{c}{ t[s]} & &
\multicolumn{1}{c}{cut [$\%$]} & \multicolumn{1}{c}{ t[s]} & & \multicolumn{1}{c}{cut [$\%$]} & \multicolumn{1}{c}{ t[s]} \\
\hline
2      &  & \numprint{1578.0}  &  & \numprint{4.1}  &  & \numprint{-4.27} & 15.2  &  & \numprint{-5.64} & 16.8 &  & \numprint{-0.16} & 0.9 &  & \numprint{+9.33}  & 0.2	\\
4      &  & \numprint{3349.2}  &  & \numprint{5.0}  &  & \numprint{-6.82} & 19.0  &  & \numprint{-6.16} & 30.9 &  & \numprint{+2.25} & 1.7 &  & \numprint{+12.09} & 0.4	\\
8      &  & \numprint{5215.2}  &  & \numprint{6.8}  &  & \numprint{-5.92} & 23.5  &  & \numprint{-4.37} & 42.9 &  & \numprint{+3.25} & 2.5 &  & \numprint{+11.77} & 0.5	\\
16     &  & \numprint{7655.8}  &  & \numprint{10.3} &  & \numprint{-4.26} & 31.6  &  & \numprint{-2.60} & 54.8 &  & \numprint{+2.61} & 3.2 &  & \numprint{+9.62}  & 0.6	\\
32     &  & \numprint{10649.4} &  & \numprint{16.7} &  & \numprint{-2.69} & 46.8  &  & \numprint{-1.73} & 66.4 &  & \numprint{+2.78} & 4.1 &  & \numprint{+8.27}  & 0.8	\\
64     &  & \numprint{14322.2} &  & \numprint{28.0} &  & \numprint{-0.76} & 72.9  &  & \numprint{-0.44} & 77.2 &  & \numprint{+3.57} & 4.8 &  & \numprint{+8.29}  & 0.9	\\
128    &  & \numprint{18316.5} &  & \numprint{42.5} &  & \numprint{+2.48} & 106.6 &  & \numprint{+0.72} & 88.9 &  & \numprint{+3.21} & 5.6 &  & \numprint{+7.22}  & 1.0	\\
\hline
avg    &  & \numprint{6673.5} &  & \numprint{11.6} &  & \numprint{-3.22} & 36.0  &  & \numprint{-2.92} & 47.6 &  & \numprint{+2.49} & 2.8 &  & \numprint{+9.50}  & 0.6	\\

\end{tabular}

\vspace{1cm}

\begin{tabular}{rc|rcrc|rrc|rcrc|rrc|rr}
\multirow{2}{*}{k} && \multicolumn{3}{c}{\textbf{FastV}}                 & & \multicolumn{2}{c}{hMetis-K} & &  \multicolumn{3}{c}{hMetis-R} & & \multicolumn{2}{c}{PaToH-Q} &  &\multicolumn{2}{c}{PaToH-D} \\
    && avg. cut  &&  t[s] & &\multicolumn{1}{c}{cut [$\%$]} & \multicolumn{1}{c}{ t[s]} & & \multicolumn{1}{c}{cut [$\%$]} && \multicolumn{1}{c}{ t[s]} & &
\multicolumn{1}{c}{cut [$\%$]} & \multicolumn{1}{c}{ t[s]} & & \multicolumn{1}{c}{cut [$\%$]} & \multicolumn{1}{c}{ t[s]} \\
\hline
2      &  & \numprint{4573.2}   &  & \numprint{225.8} &  & \numprint{-5.42} & 331.7 &  & \numprint{-10.98} && \numprint{350.5}  &  & \numprint{-7.42} & 15.4 &  & \numprint{-4.98} & 5.0	\\
4      &  & \numprint{10012.8}  &  & \numprint{233.9} &  & \numprint{-3.21} & 371.5 &  & \numprint{-9.29}  && \numprint{650.1}  &  & \numprint{-2.61} & 30.0 &  & \numprint{-1.45} & 9.2	\\
8      &  & \numprint{18887.7}  &  & \numprint{247.0} &  & \numprint{-4.41} & 403.1 &  & \numprint{-9.26}  && \numprint{914.2}  &  & \numprint{-5.78} & 44.3 &  & \numprint{-1.19} & 13.1	\\
16     &  & \numprint{31280.3}  &  & \numprint{269.6} &  & \numprint{-2.37} & 438.5 &  & \numprint{-7.89}  && \numprint{1143.5} &  & \numprint{-4.26} & 58.3 &  & \numprint{+0.86} & 16.7	\\
32     &  & \numprint{49073.0}  &  & \numprint{300.6} &  & \numprint{-1.56} & 484.7 &  & \numprint{-7.39}  && \numprint{1352.9} &  & \numprint{-3.62} & 72.7 &  & \numprint{+0.57} & 20.3	\\
64     &  & \numprint{77031.8}  &  & \numprint{343.0} &  & \numprint{-0.50} & 547.7 &  & \numprint{-7.14}  && \numprint{1552.8} &  & \numprint{-3.71} & 83.5 &  & \numprint{+2.28} & 23.8	\\
128    &  & \numprint{114556.2} &  & \numprint{398.6} &  & \numprint{-0.83} & 638.0 &  & \numprint{-5.59}  && \numprint{1733.5} &  & \numprint{-2.18} & 99.1 &  & \numprint{+2.83} & 27.3	\\
\hline
avg    &  & \numprint{27440.7}  &  & \numprint{282.8} &  & \numprint{-2.63} & 449.2 &  & \numprint{-8.23}  && \numprint{979.9}  &  & \numprint{-4.24} & 49.4 &  & \numprint{-0.19} & 14.5	\\
\end{tabular}
 \end{table}
\clearpage

\section{Conclusions and Future Work}  \label{Conclusions}
We presented the $n$-level direct $k$-way hypergraph partitioning framework KaHyPar.
Using a highly localized version of $k$-way FM, our algorithm produces better partitions
than hMetis and PaToH on $73\%$ of the VLSI instances and $71\%$ of the sparse matrix instances. 
Our greedy algorithm based on size-constrained label propagation gives better quality than PaToH on VLSI instances 
while still being several times faster than hMetis.

Motivated by the effectiveness of hMetis-R, evaluating recursive bisection in the context of $n$-level partitioning
seems to be a promising area of future research. Having both a direct $k$-way and a recursive bisection $n$-level 
partitioner, KaHyPar could then be embedded into an evolutionary framework which combines both approaches to find better solutions~\cite{SandersS12distributed}.

Throughout local search, a lot of moves have gain zero.
Integrating the concept of higher-level gains as described by
\cite{LockedNets,HypergraphKFM} therefore would be a promising approach to give these moves more
meaning. 
The running time of our $k$-way local search could be improved by developing an adaptive
stopping rule as in \cite{nGP} that is able to model zero-gain moves and stops local search if further improvement becomes unlikely.
Having shown that our localized direct $k$-way local search algorithm is able to optimize the total cut size, future work
could also look at different partitioning objectives that rely on a global view of the problem, like the $(\lambda-1)$ or \emph{sum-of-external-degrees}
metric~\cite{hMetisKway}.

{
\bibliographystyle{plain}
\bibliography{library,refs-parco}
}

\appendix
\newpage
\section{Benchmark Instances}
\begin{table}[!b]
\centering
   \caption{Properties of our benchmark set containing medium-sized instances (top) and large instances (bottom). Tables are split into two groups: VLSI instances and sparse matrix instances. Within each group, the hypergraphs are sorted by size.}
\label{tbl:mediuminstances}
\begin{tabular}{l|rcrcr|rrr|rrr}
\multirow{2}{*}{Hypergraph} & \multicolumn{1}{c}{\multirow{2}{*}{$|V|$}} && \multicolumn{1}{c}{\multirow{2}{*}{$|E|$}}  && \multicolumn{1}{c|}{\multirow{2}{*}{$|pins|$}}  & \multicolumn{3}{c|}{$d(v)$} & \multicolumn{3}{c}{$|e|$} \\
 & && && & \multicolumn{1}{c}{min} & \multicolumn{1}{c}{avg} & \multicolumn{1}{c|}{max} & \multicolumn{1}{c}{min} & \multicolumn{1}{c}{avg}& \multicolumn{1}{c}{max} \\
\hline
ibm01 &  \numprint{12752} &&  \numprint{14111} &&  \numprint{ 50566} & 1 & 3.97 &   39 & 2 & 3.58 &  42 \\
ibm02 &  \numprint{19601} &&  \numprint{19584} &&  \numprint{ 81199} & 1 & 4.14 &   69 & 2 & 4.15 & 134 \\
ibm03 &  \numprint{23136} &&  \numprint{27401} &&  \numprint{ 93573} & 1 & 4.04 &  100 & 2 & 3.41 &  55 \\
ibm04 &  \numprint{27507} &&  \numprint{31970} &&  \numprint{105859} & 1 & 3.85 &  526 & 2 & 3.31 &  46 \\
ibm05 &  \numprint{29347} &&  \numprint{28446} &&  \numprint{126308} & 1 & 4.30 &    9 & 2 & 4.44 &  17 \\
ibm06 &  \numprint{32498} &&  \numprint{34826} &&  \numprint{128182} & 1 & 3.94 &   91 & 2 & 3.68 &  35 \\
ibm07 &  \numprint{45926} &&  \numprint{48117} &&  \numprint{175639} & 1 & 3.82 &   98 & 2 & 3.65 &  25 \\
\hline
vibrobox &  \numprint{12328} &&  \numprint{12328} &&  \numprint{ 342828} &  9 & 27.81 & 121 &  9 & 27.81 & 121 \\
bcsstk29 &  \numprint{13992} &&  \numprint{13992} &&  \numprint{ 619488} &  5 & 44.27 &  71 &  5 & 44.27 &  71 \\
memplus  &  \numprint{17758} &&  \numprint{17758} &&  \numprint{ 126150} &  2 &  7.10 & 574 &  2 &  7.10 & 574 \\
bcsstk30 &  \numprint{28924} &&  \numprint{28924} &&  \numprint{2043492} &  4 & 70.65 & 219 &  4 & 70.65 & 219 \\
bcsstk31 &  \numprint{35588} &&  \numprint{35588} &&  \numprint{1181416} &  2 & 33.20 & 189 &  2 & 33.20 & 189 \\
bcsstk32 &  \numprint{44609} &&  \numprint{44609} &&  \numprint{2014701} &  2 & 45.16 & 216 &  2 & 45.16 & 216 \\
\end{tabular}

\vspace{.3cm}

\begin{tabular}{l|rcrcr|rrr|rrr}
\multirow{2}{*}{Hypergraph} & \multicolumn{1}{c}{\multirow{2}{*}{$|V|$}} && \multicolumn{1}{c}{\multirow{2}{*}{$|E|$}}  && \multicolumn{1}{c|}{\multirow{2}{*}{$|pins|$}}  & \multicolumn{3}{c|}{$d(v)$} & \multicolumn{3}{c}{$|e|$} \\
 & && && & \multicolumn{1}{c}{min} & \multicolumn{1}{c}{avg} & \multicolumn{1}{c|}{max} & \multicolumn{1}{c}{min} & \multicolumn{1}{c}{avg}& \multicolumn{1}{c}{max} \\
\hline
ibm08 &  \numprint{ 51309} &&  \numprint{ 50513} &&   \numprint{204890} &  1 & 3.99 & 1165 & 2 & 4.06 &  75 \\
ibm09 &  \numprint{ 53395} &&  \numprint{ 60902} &&   \numprint{222088} &  1 & 4.16 &  173 & 2 & 3.65 &  39 \\
ibm10 &  \numprint{ 69429} &&  \numprint{ 75196} &&   \numprint{297567} &  1 & 4.29 &  137 & 2 & 3.96 &  41 \\
ibm11 &  \numprint{ 70558} &&  \numprint{ 81454} &&   \numprint{280786} &  1 & 3.98 &  174 & 2 & 3.45 &  24 \\
ibm12 &  \numprint{ 71076} &&  \numprint{ 77240} &&   \numprint{317760} &  1 & 4.47 &  473 & 2 & 4.11 &  28 \\
ibm13 &  \numprint{ 84199} &&  \numprint{ 99666} &&   \numprint{357075} &  1 & 4.24 &  180 & 2 & 3.58 &  24 \\
\end{tabular}
\end{table}

\clearpage

\changepage{}{3.5cm}{-2.75cm}{-1cm}{}{5.5cm}{}{}{}
\pagestyle{empty} 
\begin{landscape}
\section{Detailed Results}
\tiny
\setlength{\tabcolsep}{1pt}
\begin{longtable}{l|r|rrr|rrr|rrr|rrr|rrr|rrr|rrr|rrr}
\caption{\normalsize Detailed results per instance. The best cut is highlighted.} \label{tbl:detailedResults}\\
  &                        & \multicolumn{3}{c|}{KaHyPar-Strong} & \multicolumn{3}{c|}{KaHyPar-StrongV} & \multicolumn{3}{c|}{KaHyPar-Fast} & \multicolumn{3}{c|}{KaHyPar-FastV} & \multicolumn{3}{c|}{hMetis-K} & \multicolumn{3}{c|}{hMetis-R} &\multicolumn{3}{c|}{PaToH-Q} &\multicolumn{3}{c}{PaToH-D}  \\
H & k                      & Best & Avg.&  t[s] & Best & Avg.     & t[s] & Best & Avg.   & t[s] & Best & Avg. & t[s] & Best & Avg. & t[s] & Best & Avg.   & t[s] & Best & Avg.  & t[s] & Best & Avg. & t[s]  \\ \hline
\endfirsthead
  &                        & \multicolumn{3}{c|}{KaHyPar-Strong} & \multicolumn{3}{c|}{KaHyPar-StrongV} & \multicolumn{3}{c|}{KaHyPar-Fast} & \multicolumn{3}{c|}{KaHyPar-FastV} & \multicolumn{3}{c|}{hMetis-K} & \multicolumn{3}{c|}{hMetis-R} & \multicolumn{3}{c|}{PaToH-Q} & \multicolumn{3}{c}{PaToH-D}  \\
H & k                      & min & avg&  T[s] & min & avg     & T[s] & min & avg   & T[S] & min & avg & T[s] & min & avg & T[s] & min & avg   & T[s] & min & avg  & T[s] & min & avg & T[s]  \\ \hline
\endhead
\multirow{7}{*}{\rotatebox[origin=c]{90}{ibm01}}       & \numprint{2}   & \textbf{\numprint{203}}   & \numprint{243.3}           & \numprint{1.54}      & \textbf{\numprint{203}}    & \numprint{241.3}           & \numprint{3.41}      & \numprint{266}          & \numprint{274.0}     & \numprint{0.22}    & \numprint{261}              & \numprint{266.0}     & \numprint{0.38}    & \textbf{\numprint{203}}   & \numprint{206.3}           & \numprint{0.86}     & \textbf{\numprint{203}}     & \numprint{203.1}           & \numprint{1.17}     & \numprint{252}             & \numprint{252}             & \numprint{0.15}    & \numprint{265}             & \numprint{290.3}     & \numprint{0.03}\\
                                                       & \numprint{4}   & \numprint{583}            & \numprint{600.0}           & \numprint{3.89}      & \numprint{579}             & \numprint{596.8}           & \numprint{9.42}      & \numprint{600}          & \numprint{622.1}     & \numprint{0.48}    & \numprint{590}              & \numprint{610.5}     & \numprint{0.63}    & \textbf{\numprint{495}}   & \numprint{520.4}           & \numprint{1.48}     & \numprint{535}              & \numprint{537.2}           & \numprint{2.50}     & \numprint{640}             & \numprint{640}             & \numprint{0.23}    & \numprint{546}             & \numprint{656.5}     & \numprint{0.04}\\
\hline
\end{longtable}
\end{landscape}
\end{document}